# Capillary-driven biaxial planar and homeotropic nematization of hard cylinders

Roohollah Aliabadi[a,*], Soudabe Nasirimoghadam[a], Henricus Herman Wensink[b]

[a] *Physics Department, Sirjan University of Technology, Sirjan 78137, Iran*

[b] *Laboratoire de Physique des Solides - UMR 8502, CNRS, Université Paris-Saclay, 91405 Orsay, France*

**Abstract**

We use the Parsons–Lee modification of Onsager's second virial theory within the restricted orientation (Zwanzig) approximation to analyse the phase behaviour of hard cylindrical rods confined in narrow pores. Depending on the wall-to-wall separation we predict a number of distinctly different surface-generated nematic phases, including a biaxial planar nematic with variable number of layers, a monolayer homeotropic, and a hybrid T-type structure (a planar layer combined with a homeotropic one). For narrow pores, we find evidence of two types of second-order uniaxial-biaxial transitions depending on the aspect ratio of the particles. More specifically, we observe a continuous cross-over from $n$ to $n+1$ layers each with a distinct planar anchoring symmetry as well as first-order transitions from planar to homeotropic surface anchoring. Contrary to the previously studied case of parallelepipeds we find that the surface anchoring transition from planar to homeotropic symmetry occurs at much lower overall rod packing fractions. This renders the observation of homeotropic capillary nematics much more realistic in experimental systems of strongly confined anisotropic colloids. Unlike confined parallelepipeds, cylindrical rods gradually increase the number of the nematic planar layers (without any phase transitions). However, a weak first order transition was observed between two planar structures with $n$ and $n+1$ layers in wide pores and longer rods. In addition, the cylindrical rods exhibit a first-order transition from the homeotropic structure to the uniaxial (or biaxial) T phase that has not been observed in confined hard parallelepipeds. We further demonstrate a reentrant uniaxial-biaxial-uniaxial-biaxal phase sequence for confined cylinders at small aspect ratio. Our results also clearly demonstrate that stable T-type surface ordering is a subtle capillary effect that only becomes manifest in sufficiently narrow pores away from the 2D bulk limit.

**Introduction**

Although the discovery of liquid crystals (LCs) dates back more than a century ago, LCs are still studied eagerly in an effort to search for new materials that meet the constantly growing need of applications in biosensors and organic transistors [1,2] and to further our scientific understanding self-organization of soft condensed matter in the presence of external fields such as geometric confinement [3–5]. As a benchmark model colloidal particles with a simple non-isotropic shape, such as rods and discs, have been analyzed extensively in theory [6–9], experiment [10,11] and by computer simulation [12–14].

Onsager (1949) [15,16] argued that the key factor in the formation of liquid crystalline structures is the anisotropic shape of the particles for which simple geometrical shapes are usually considered, such as ellipsoids [13,17] cylindrical rods [18], cut spheres [19], or parallelepipeds [20,21].

In many technological applications involving non-isotropic colloidal particles the influence of system boundaries is ubiquitous. It is therefore important to understand self-organization of colloidal rods under the condition of strong geometric confinement. In particular, confining anisotropic particles in nanopores can provides a way of controlling their phase behavior and stabilize novel structures arising from the competition between intrinsic and extrinsic particle interactions [22]. Planar (main rod axes parallel to the wall), homeotropic (rod axes perpendicular to the wall), tilted (rod axes making an oblique angle with the wall), and more complicated phases such as the hybrid T structure (a planar layer combined with a homeotropic one) are examples of capillary states that can form between two parallel surfaces. However, according to simulations studies, the prevailing anchoring symmetry seems to be planar-type [12,23]. The hybrid T-type structure was only found to emerge for very weakly anisotropic parallelepiped through a first order phase transition from the planar structures [24].

---

[*] aliabadi313@gmail.com

Schoen and coworkers simulated the effects of different wall-particle interactions on the stratification (layering) of thin confined films with various molecular fluids [25–28]. The external field applied by the walls on the film forced the particles to arrange in particular directions reflecting the molecular structures in the local density. It was demonstrated that the layering phenomena were not very sensitive to the details of the particle–wall interaction, *i.e.*, simple systems such as hard spheres constrained between hard walls [29] or soft Lenard–Jones atoms confined between molecularly structured or unstructured walls [30–33] exhibit similar types of stratification. The main finding is that if the walls lack molecular structure, they cannot induce the formation of solid films. Hence, the film remains fluid, and new layers are formed sequentially in a continuous manner without changing the basic fluid symmetry of the film [25]. In 2018, a first-order nematic planar layering transition was reported for the first time in hard parallelepipeds confined between two unstructured hard walls in [6]. This study reports a (sequence of) nematic planar layering transitions in sufficiently narrow pores from a nematic planar with $n$ layers to a nematic planar with $n+1$ layers terminating at a critical pore width which is sufficiently wide.

Studies on confined hard rods confirm that elongated particles tend to anchor in a planar fashion at the walls whereas strong confinement promotes the formation of a nematic phase through capillary nematization involving complete wetting [20,34]. The presence of a surface also stabilizes high-density phases such as smectics and solids. For instance, de las Heras *et al.* [35] reported a layering transition in the smectic structure for confined spherocylinders. They indicated the existence of a first-order layering transition from the smectic phase with $n$ layers to $n+1$ smectic layers in wide pores with the two phases meeting a nematic phase at a triple point. For sufficiently narrow pores it was demonstrated that the triple point vanishes with the smectic layering transition terminating at a critical point.

The present theoretical study focuses on the occurrence of biaxial phases and layered nematic structures of various anchoring symmetries using the Parsons–Lee (PL) theory for cylindrical rods. The main rationale behind the latter is that cylinders are usually considered a more accurate representation of colloidal shapes found in experiment than parallelepipeds [6]. In our treatment, the main rod vectors are restricted to point along one of the Cartesian axes, as per the Zwanzig [36] or the restricted orientation model (ROM) which greatly reduces the computational expense of our theory.

Despite the simplicity of the ROM, the approximation seems to capture the basic physics of confined particles with predictions being qualitatively similar to those for freely rotating particles. The studies in Refs. [20, 34, 37,38] confirm the qualitative validity of most phase behaviours emerging from Zwanzig theory for a wide class of hard-rod systems. However, for bulk system it is known that ROM can indeed lead to qualitatively different results. For instance, density functional methods based on ROM tend to overestimate the phase transition densities and predict artificial aligned isotropic or nematic phase in bulk [39]. For example, for certain values of aspect ratios a peculiar bulk smectic phase [40] was found. If the restriction of orientations is removed, several simulation studies of hard prolate particles show that this peculiar smectic is never stable. Thus, its stability reported in Ref. [40] is obviously an artifact of the restriction on orientations. However, for strongly confined particles one can argue that the ROM approximation is much better justified give that particle orientations are naturally restricted by the walls as well as by the end effects imparted by the cylinders in particular in the regime of small aspect ratios that we consider in this study. Clearly a definitive assessment of our theoretical model can only be given through simulating confined fluids of hard cylinders using freely-rotating hard cylinders, which we do not pursue here.

**Theory**

We consider the combined positional and orientational ordering of cylindrical hard rods with a length $L$ and cross-section $D$ confined between two identical, parallel hard structureless walls normal to the $z$-axis with wall-to-wall separation $H$. The origin of the Cartesian system coincides with one of the walls. All interactions in our system are hard; therefore, the particles are not permitted to penetrate each other or the wall, so that all structures are purely entropy-driven. To examine the structures emerging from the confined hard cylinders, PL theory [41, 42] was applied to systems in which $H > L$. Although the Onsager–Parsons–Lee theory is approximate, it accurately reproduces the simulation results for liquid crystals composed of weakly anisotropic particles in the quasi-one-dimensional limit [43,44]. A three-state restricted orientation approximation is used for the orientation of the long axis of particles (Zwanzig approximation [36]), *i.e.*, the main symmetry axes of the particles only align along one of the three Cartesian directions ($x$, $y$, and $z$). As such, the system is equivalent to a ternary mixture of hard rods each with no rotational freedom.

In our model the external potential is a function of $z$ alone, and we do not consider in-plane crystallization, for instance, mediated by a Kosterlitz–Thouless transition [45]. The local density components ($\rho_i$, $i = x, y$ and $z$) depend only on the perpendicular distance ($z$) from the walls. To determine the local density components in inhomogeneous fluids, we consider the Onsager–Parsons–Lee grand canonical free energy ($\Omega[\rho]$) as the key ingredient:

$$\frac{\beta \Omega[\rho]}{A} = \sum_{i=x,y,z} \left[ \int dz \rho_i(z) \big(\beta V^i_{ext}(z) - \beta\mu\big) + \int dz \rho_i(z) (\ln \rho_i(z) - 1) \right]$$

$$+ \frac{1}{2} c \sum_{i,j=x,y,z} \int dz_1 \rho_i(z_1) \int dz_2 A^{ij}_{exc}(z_1 - z_2) \rho_j(z_2), \qquad (1a)$$

where $\beta V^i_{ext}$ denotes the external potential for a rod with orientation $i$, and $\mu$ represents the chemical potential. Moreover, $\beta = 1/(k_B T)$ and $c = (1 - 3\eta/4)/(1 - \eta)^2$ is the Parsons–Lee pre-factor, whereas $A^{ij}_{exc}$ represent the excluded area in the $x$-$y$ plane between two cylinders with orientations $i$ and $j$.

The surface–rod interactions are different for particles aligning parallel ($x$ and $y$ orientations) and perpendicular ($z$ orientation) to the walls,

$$\beta V^x_{ext}(z) = \beta V^y_{ext}(z) = \begin{cases} 0, & D/2 \le z \le H - D/2 \\ \infty, & \text{otherwise} \end{cases} \qquad (2)$$

and

$$\beta V^z_{ext}(z) = \begin{cases} 0, & L/2 \le z \le H - L/2 \\ \infty, & \text{otherwise} \end{cases} \qquad (3)$$

In this formula, the packing fraction ($\eta$) can be obtained from the local densities through the following equation:

$$\eta = \frac{v_0}{V} \sum_{i=x,y,z} \int d\vec{r} \rho_i(\vec{r}) = \frac{v_0}{H} \left( \int_{D/2}^{H-D/2} \rho_x(z) dz + \int_{D/2}^{H-D/2} \rho_y(z) dz + \int_{L/2}^{H-L/2} \rho_z(z) dz \right), \quad (4)$$

where $v_0 = (\pi D^2 L)/4$ is the volume of a cylinder, and $V = AH$ indicates the pore volume. Moreover, $A$ denotes the wall area.

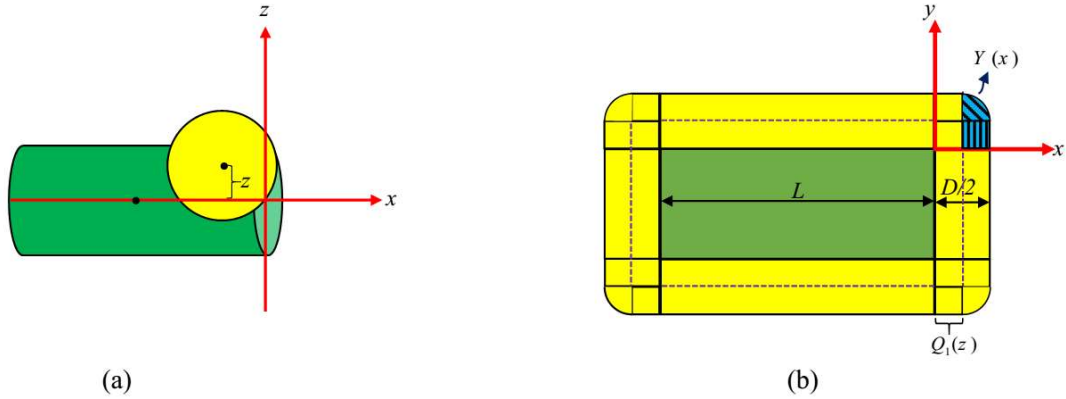

FIG. 1: Sketch of the excluded volume between two perpendicular cylinders. (a) The yellow cylinder is parallel to the *y*-axis (normal to the page) and swipes around a fixed green cylinder parallel to the *x*-axis. The two cylinders are at close contact. The excluded surface swept out in the *x-y* plane is shown in panel (b) where $Y(x)$ and $Q_1(z)$ in Eq. (5(c)) represent the excluded areas, *i.e.*, $A^{xy}_{exc}(z) = A^{yx}_{exc}(z)$ where $z$ denotes the distance of their center of masses along the *z*-axis. Note these cylinders cannot penetrate each other.

The excluded areas between two cylinders for all possible combinations of orientations in Eq. (1) are not straightforwardly computed within the Zwanzig model. For instance, Fig. 1 demonstrates the method of calculating the excluded area between two cylinders when one of them is fixed and parallel with the *x*-axis, whereas the other one is aligned with the *y*-axis moving around the first rod. The excluded areas are as follows:

$$A^{xx}_{exc}(z) = A^{yy}_{exc}(z) = 4L\sqrt{D^2 - z^2} \quad \text{for} \quad |z| \leq D \tag{5a}$$

$$A^{zz}_{exc}(z) = \pi D^2 \quad \text{for} \quad |z| \leq L/2 \tag{5b}$$

$$\begin{cases} A^{xy}_{exc}(z) = A^{yx}_{exc}(z) = 2LD + L^2 + 2DQ_1(z) + 4\int_{Q_1(z)}^{D/2} Y(x)dx & \text{for} \quad |z| \leq D/2 \\ A^{xy}_{exc}(z) = A^{yx}_{exc}(z) = L^2 + 4LQ_2(z) + 4\int_0^{Q_2(z)} Y(x)dx & \text{for} \quad D/2 \leq |z| < D \end{cases} \tag{5c}$$

where $Q_1(z) = \sqrt{\frac{D^2}{4} - z^2}$, $Q_2(z) = \sqrt{\frac{D^2}{4} - (|z| - \frac{D}{2})^2}$, and $Y(x) = \sqrt{\frac{D^2}{4} - (\sqrt{\frac{D^2}{4} - x^2} - |z|)^2}$. The integral in the top line of Eq. (5(c)) represents the blue patterned region in Fig. 1(b). The remaining excluded areas can be calculated as follows:

$$\begin{cases} A^{xz}_{exc}(z) = A^{yz}_{exc}(z) = A^{zx}_{exc}(z) = A^{zy}_{exc}(z) = D^2 + 2LD + \pi(\frac{D}{2})^2 & \text{for} \quad |z| \leq L/2 \\ A^{xz}_{exc}(z) = A^{yz}_{exc}(z) = A^{zx}_{exc}(z) = A^{zy}_{exc}(z) = DD_1 + \pi(\frac{D}{2})^2 + L(D + D_1) & \text{for} \quad L/2 \leq |z| < (L+D)/2 \end{cases} \tag{5d}$$

where $D_1 = \sqrt{D^2 - 4(|z| - L/2)^2}$.

The free energy Eq. (1) has to be minimized with respect to all the density components to determine the equilibrium of local densities. The subsequent coupled equations were obtained through functional differentiation:

$$\ln \rho_k(z) - \beta\mu + \beta V_{ext}^k(z) + \frac{1}{2}\frac{dc}{d\eta}\frac{\delta\eta}{\delta\rho_k(z)}\sum_{i,j=x,y,z}\int dz_1 \rho_i(z_1)\int dz_2 A_{exc}^{ij}(z_1-z_2)\rho_j(z_2)$$

$$+\frac{(1-3\eta/4)}{(1-\eta)^2}\sum_{i=x,y,z}\int dz_1 A_{exc}^{ik}(z-z_1)\rho_i(z_1) = 0, \quad (6)$$

where $k = x, y, z$. In Eq. (6), $dc/d\eta = (5-3\eta)/4(1-\eta)^3$, and $\delta\eta/\delta\rho_k(z) = v_0/H$ was replaced. Therefore, the local density components can be rearranged into the following expression:

$$\rho_k(z) = \exp\left[\beta\mu - \frac{v_0(5-3\eta)}{8H(1-\eta)^3}\sum_{i,j=x,y,z}\int dz_1 \rho_i(z_1)\int dz_2 A_{exc}^{ij}(z_1-z_2)\rho_j(z_2)\right] \times$$

$$\exp\left[-\frac{(1-3\eta/4)}{(1-\eta)^2}\sum_{i=x,y,z}\int dz_1 A_{exc}^{ik}(z-z_1)\rho_i(z_1) - \beta V_{ext}^k(z)\right] \quad (7)$$

The three self-consistent equations must be solved iteratively for given values of $L/D$, $H/D$, and $\beta\mu$. Since it is preferable to work with an overall rod packing fractions $\eta$ rather than chemical potential $\beta\mu$, the local density equations are determined in terms of $\eta$ by substituting Eq. (7) in Eq. (4), extracting $\exp[\beta\mu]$, replacing it in Eq. (7), and finally inserting it in Eq. (4). This gives:

$$\rho_k(z) = \frac{H\eta \exp\left[-\beta V_{ext}^k(z)-\frac{(1-3\eta/4)}{(1-\eta)^2}\sum_{i=x,y,z}\int dz_1 A_{exc}^{ik}(z-z_1)\rho_i(z_1)\right]}{v_0 \sum_{q=x,y,z}\int dz_1 \exp\left[-\beta V_{ext}^q(z_1)-\frac{(1-3\eta/4)}{(1-\eta)^2}\sum_{i=x,y,z}\int dz_2 A_{exc}^{iq}(z_1-z_2)\rho_i(z_2)\right]} \quad (8)$$

We present the results in dimensionless units where $D$ is defined as our unit of distance, i.e., $\rho^* = \rho D^3$ and $z^* = z/D$. To determine the equilibrium local density components, Eq. (8) must be solved numerically after the discretization of $z$ positions of particles through equidistant grid points ($\Delta z = H/m, z_i = i\Delta z, i = 0,\ldots,m$) by using Picard's iteration procedure using a linear combination rule that mixes the results of successive iterations. The criterion $\frac{1}{m+1}\sum_{i=0}^{m}\sum_{k=x,y,z}|\rho_k^{l+1}(z_i) - \rho_k^l(z_i)| < 10^{-9}$ was employed to control the convergency of the consecutive iterative steps. In this equation, $l$ is an iteration step where the convergence occurs. A grid size of $\Delta z = 0.01$ was used in case the phase separation is strong, while finer discretizations $\Delta z = 0.001$ or $0.0001$ were employed in case the coexisting densities were very close. In the iteration procedure, it is essential to select the correct initial value to obtain reliable solutions for each phase. For this purpose, $\rho_x > 0, \rho_y > 0$ and $\rho_z = 0$ were selected for the planar structure. Although the $x$ and $y$ orientations are equivalent, the trial distribution functions for the local densities are selected so that they may result in $\rho_x \geq \rho_y$. For the homeotropic phase, these initial values $\rho_x = \rho_y = 0$ and $\rho_z > 0$ are selected. In addition, the trapezoidal quadrature method is employed to calculate the integrals. The reliability of the discretization approximation and the convergence of successive iteration applied to the Euler–Lagrange integral equations were discussed by Herzfeld et al. [46].

Discontinuous phase transitions are determined from the cross point of two different solutions of Eq. (8) in the $\beta\mu - \beta\Omega/A$ plane. This is equivalent to the equality of pressure and the chemical potential of two phases in coexistence. Moreover, it is possible to find several different phases at a certain packing fraction; however, the stable phase in the system is always the one that has the lowest free energy.

**Results and Discussion**

This section concentrates on the phase structures of confined hard cylindrical rods with $H - 2D < L < H$, where only the planar (P) state with a maximum of $n+1$ layers, a homeotropic (H) layer and one T-form layer of cylinders can accommodate in the pores. The pore size we consider are small enough to guarantee that no uniform nematic fluid can form within the central region of the pores at intermediate and high densities. For instance, when $H < 3L$, no capillary nematization can be detected for confined hard rods [34]. We stress that, in the general case, a first order isotropic-nematic transition (capillary nematization) may however, still show up even in very small pores by modifying the sharpness of the pair interactions [47]. We consider a range of aspect ratios between $1.80 \leq L/D < 12.00$ and seven different wall-to-wall separations, i.e., $H/D = 3.1, 3.3, 3.5, 4.5, 8.0, 10.0$, and 12. The maximum allowable planar layers is either $n + 1 = [H/D] - 1$ for an integer $H/D$ or $n + 1 = [H/D]$ for a non-integer $H/D$, in which $[H/D]$ denotes the integer part of $H/D$.

Fig. 2 depicts a schematic diagram of the examined phase structures for $H/D = 3.1$ and $L/D = 2.00$. This figure is a comprehensive representation of all possible phases which may appear upon increasing $\eta$. In Fig. 3(a) we show the phase diagram for $H/D = 3.1$ in terms of the packing fraction $\eta$ versus the particle aspect ratio ($1.80 \leq L/D < 3.10$). Note that these particles are non-nematogenic in bulk. Previous simulation studies of hard rods demonstrated that confining weakly anisotropic particles in narrow pores stabilizes the formation of liquid crystalline structures even when they are unable to form such mesophases in bulk [48,49].

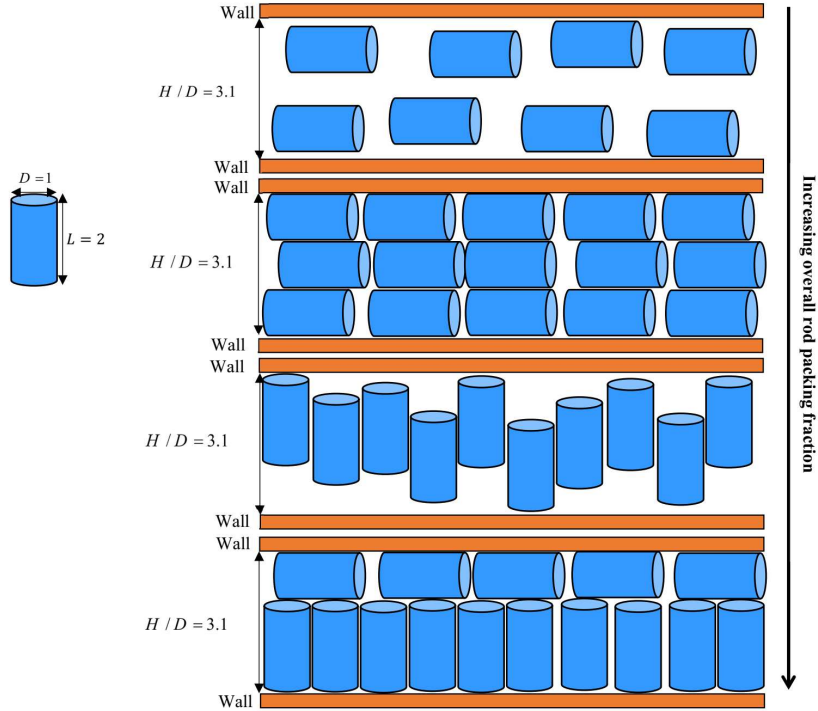

FIG. 2: A schematic representation of the possible structures of confined hard cylinders by increasing the overall rod packing fraction for $H/D = 3.1$ and $L/D = 2.00$. The panels indicate the following structures from top to bottom: a nematic ordering of the particles with two planar layers, three planar layers, a single homeotropic layer, and a T-type hybrid planar-homeotropic layer, respectively.

Figs. 3 (b)–(f) depict the density distributions at five different packing fractions, in which panels can be recognized by different symbols in Fig. 3 (a). These curves are based on numerical solution of the Euler–Lagrange equation (Eq. (8)). The different phases are summarized as below:

(b): Uniaxial structure with three planar layers at $(L/D, \eta) = (2.20, 0.39)$ where the layer in the central region of the pore is very weak. Here, determination of the number of planar layers is not straightforward. For instance, in order to judge whether a structure has two or three planar layers in Fig. 3 (b), we consider the very weakly developed planar layer in the middle of the pore as the third layer if $\rho_x(z = H/2) \neq \rho_x(z = H/2 \pm \Delta z)$ where $\Delta z$ is the grid size used in the numerical calculations.

(c): Biaxial structure with three planar layers at $(L/D, \eta) = (2.55, 0.39)$;

(d): Homeotropic monolayer at $(L/D, \eta) = (2.55, 0.47)$;

(e): Uniaxial T phase at $(L/D, \eta) = (1.90, 0.65)$ with a homeotropic layer at one wall and a uniaxial planar layer at the opposing one;

(f): Biaxial T phase at $(L/D, \eta) = (1.90, 0.65)$ with a homeotropic layer at one wall and a biaxial planar layer at the opposing one.

In Fig. 3(a), the blue solid line and the dashed red line present a second-order transition from the uniaxial planar (UP) to the biaxial planar (BP) and from a uniaxial (UT) to a biaxial (BT) T-type phase, respectively. The formation of a biaxial nematic wetting layer has been demonstrated in a number of studies [34,50]. The resulting first-order phase transition from P to H is highlighted in a green area and patterned with vertical lines within the range of $1.90 \leq \frac{L}{D} < 3.10$. The T structure terminates at two critical aspect ratios $\frac{L}{D} \approx 1.83$ and $2.09$, occurs through a first order transition from 3UPL, 3BPL, and 1HL. According to Fig. 3(a), there are seven different concentration-driven phase sequences each corresponding to a different range of aspect ratios:

- for $2.81 \leq L/D < 3.10$:
  $$2\text{UPL} \xrightarrow{\text{second order transition}} 2\text{BPL} \xrightarrow{\text{gradually}} 3\text{BPL} \xrightarrow{\text{first order transition}} 1\text{HL}$$

- for $2.44 \leq L/D < 2.81$:
  $$2\text{UPL} \xrightarrow{\text{gradually}} 3\text{UPL} \xrightarrow{\text{second order transition}} 3\text{BPL} \xrightarrow{\text{first order transition}} 1\text{HL}$$

- for $2.09 \leq L/D < 2.44$:
  $$2\text{UPL} \xrightarrow{\text{gradually}} 3\text{UPL} \xrightarrow{\text{first order transition}} 1\text{HL}$$

- for $2.05 \leq L/D < 2.09$:
  $$2\text{UPL} \xrightarrow{\text{gradually}} 3\text{UPL} \xrightarrow{\text{first order transition}} 1\text{HL} \xrightarrow{\text{first order transition}} 1\text{BT}$$

- for $1.94 \leq \frac{L}{D} < 2.05$:
  $$2\text{UPL} \xrightarrow{\text{gradually}} 3\text{UPL} \xrightarrow{\text{first order transition}} 1\text{HL} \xrightarrow{\text{first order transition}} 1\text{UT} \xrightarrow{\text{second order transition}} 1\text{BT}$$

- for $1.90 \leq \frac{L}{D} < 1.94$:
  $$2\text{UPL} \xrightarrow{\text{gradually}} 3\text{UPL} \xrightarrow{\text{second order transition}} 3\text{BPL} \xrightarrow{\text{first order transition}} 1\text{HL} \xrightarrow{\text{first order transition}} 1\text{UT}$$
  $$\xrightarrow{\text{second order transition}} 1\text{BT}$$

- for $1.86 \leq \frac{L}{D} < 1.90$:
  $$2\text{UPL} \xrightarrow{\text{gradually}} 3\text{UPL} \xrightarrow{\text{second order transition}} 3\text{BPL} \xrightarrow{\text{first order transition}} 1\text{UT}$$
  $$\xrightarrow{\text{second order transition}} 1\text{BT}$$

- for $1.83 \leq \frac{L}{D} < 1.86$:
  $$2\text{UPL} \xrightarrow{\text{gradually}} 3\text{UPL} \xrightarrow{\text{first order transition}} 1\text{UT} \xrightarrow{\text{second order transition}} 1\text{BT}$$

For example, 2UPL, 3BPL, and 1HL respectively denote a structure with two planar-layers each with uniaxial symmetry ($\rho_x = \rho_y$), three biaxial planar layers ($\rho_x \neq \rho_y$) including very weakly developed homeotropic layer in which the particles are oriented normal to the walls, and a homeotropic monolayer phase.

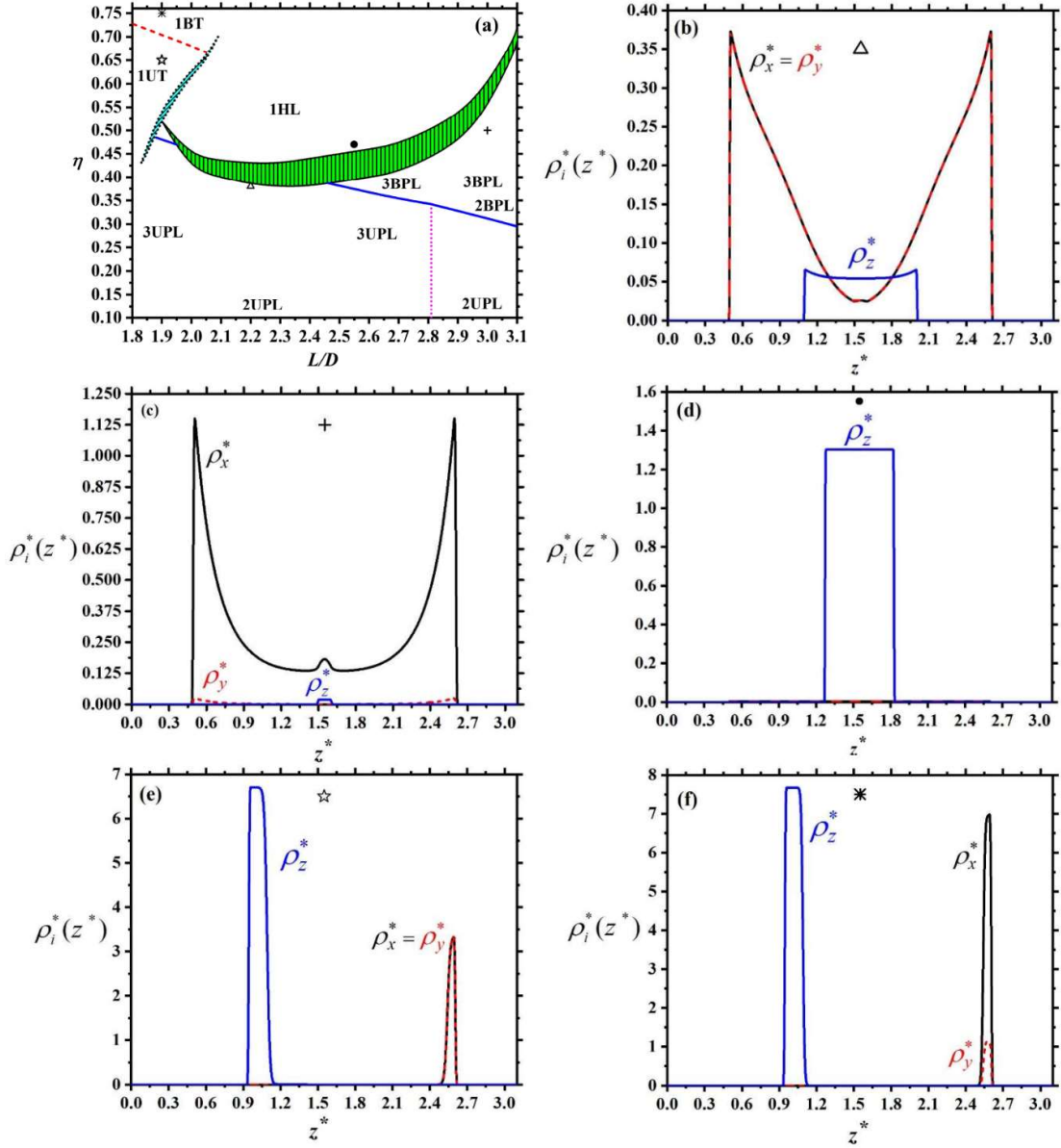

FIG. 3: (a) Phase diagram of $H/D = 3.1$ when $1.80 \leq L/D < 3.10$. Density profiles at (b) $(L/D, \eta) = (2.20, 0.39)$ where the phase is 3UPL with a very weak planar layer in the middle of the pore (c) $(L/D, \eta) = (3.00, 0.50)$ in which the structure is 3BPL (d) $(L/D, \eta) = (2.55, 0.47)$ where the state is 1HL (e) $(L/D, \eta) = (1.90, 0.65)$, whereas the phase is 1UT, and (f) $(L/D, \eta) = (1.90, 0.75)$ where the structure is 1BT.

Regarding the presence of two kinds of UP-BP transitions, since the particles with $L/D < 2.81$ are packed very well within the three-layer uniaxial planar state in the pore, it is possible to stabilize the uniaxial planar layer at higher densities and have a larger number of uniaxial planar layers. Hence, there is a second-order 3UPL-3BPL transition for $L/D < 2.81$ instead of 2UPL-2BPL. The vertical dotted pink line in Fig. 3(a) at $\frac{L}{D} = 2.81$ indicates where the biaxial planar structure emerges for a film with 3 planar layers for $L/D < 2.81$.

The excess free energy (or the last term) in Eq. (1) is responsible for stabilizing all the liquid crystalline structures, for it includes a minimum for that phase which has the minimum excluded volume between two particles. Therefore, the phase changes arise from the competition between this term and the second integral in the first term (*i.e.*, ideal part) to enhance the accessible space for the rods and give the lowest free energy. Unlike the excess part of the free energy, the ideal term prefers further disordering in the positions and orientations of the molecules. Consequently, the particles decrease the excluded areas between themselves near the walls with strong adsorption at the confining surfaces even at very low densities by selecting planar ordering. The UP-BP and UT–BT transition densities depend on the elongation of the particles ($L/D$) as for more anisotropic rods the excluded area between them expands. Thus, the second term of the free energy increases; hence, these transitions occur at lower densities for more elongated rods (refer to Fig.3 (a) and 4). In addition, as the pore width increases, the absolute values of the slopes of the UP-BP and UT–BT lines decrease due to the growth of the length-to-diameter ratio of confined particles by increasing the wall-to-wall separation since $H - 2D < L < H$. The BT phase can only be stable at very high densities given that the shape anisotropy is very weak.

According to Fig. 3(a), the first-order P-H transition terminates at a critical $(L/D)_c = 1.90$, which is smaller than $\frac{L}{D} = \frac{H}{D} - 1$ that is contrary to the behavior of confined parallelepipeds [50,51]. For $1.94 \leq L/D < 2.44$, the free energy value of 1HL is lower than the free energy of 3BPL; hence, the 1HL phase preempts the 3BPL structure, and the BP phase is absent within this range. It is harder to switch larger $L/D$s from P to H, because it is more difficult to accommodate longer rods homeotropically within the pore. Therefore, the P–H transition occurs at higher densities. This can be attributed to the fact that the translational entropy decreases along the *z*-axis in the homeotropic phase as the elongation of the particles increases. Therefore, the particles prefer to stay in the planar structure rather than the homeotropic one. This is why the P phase has a widening stability region when $L/D \to 3.1$. However, it is harder to perceive the stabilization of the planar phase with respect to the homeotropic order for $L/D < 2.30$ where the coexisting densities of 3UPL and 1HL start moving up, since the translational entropy (available room) is now more dominant in the homeotropic order. Therefore, the packing entropy (excluded area) can be the main cause of this phenomenon. This is why in the range of $1.90 \leq L/D < 1.94$ the 1HL structure occurs after the occurrence of 3BPL.

The cyan island that is patterned with tilted lines and surrounded by a dotted fence indicates the coexisting region where the T structure starts to form. According to Fig. 3 (a), this phase occurs at higher densities for larger rods because it is harder to accommodate increasingly larger particles in this structure. Therefore, it occurs above the homeotropic phase for $1.90 \leq L/D \leq 2.09$. To the best of our knowledge, a first-order 1HL-T transition has not been reported previously in literature. Such a transition cannot occur for confined hard parallelepipeds because their P–H critical points take place at an $L > (H - D)$ [51]. For $1.83 \leq L/D < 1.90$, the T structure occurs from 3UPL (or 3BPL) straightly because the formation of the T state is easy for smaller rods. The first-order transition from 3UPL to 1UT terminates at a critical aspect ratio ($(\frac{L}{D})_c \approx 1.83$), and 1UT can form continuously from 3UPL for smaller particles ($\frac{L}{D} < 1.83$).

Note that a reentrant surface ordering phenomenon occurs for $1.86 \leq \frac{L}{D} < 1.90$ where a U-B-U-B phase sequence takes place with increasing density. The low-density uniaxial phase with three planar layers and the high-density uniaxial phase with T ordering are interrupted by biaxial nematic order. In case of biaxial order there is a 1UT phase between the low and high density biaxial nematic phases.

Based on Fig. 4 (a), the phase structures of $H/D = 3.3$ are nearly identical to those of $H/D = 3.1$ including a point where the biaxiality changes its nature. Such a behavior was also reported for parallelepipeds in Ref. [50] in which the 2UPL–2BPL line intersects with the coexisting region of the first-order transition of the planar ordering with two layers to the planar with three layers at $L/D \approx 2.89$ where a new biaxiality from 3UPL to 3BPL emerges above this coexisting region when a reentrant phenomenon occurs. According to Fig. 4(a), (b) and (c), the coexisting region of the T structure (patterned regions with tilted lines and highlighted with cyan color) intersects with the coexisting region of the H phase (patterned regions with vertical lines and highlighted with green color) before it reaches its critical point. The width of the coexisting T structure increases as the pore widens, because it starts from at larger aspect ratio (*i.e.*, about $H - D$) and continues to smaller particles. A reentrant phenomenon (i.e., U-B-U-B phase sequence) is also present for $\frac{H}{D} = 3.3, 3.5$ and 4.5 where the stability region for the T-phase appears to enhance somewhat with pore size.

In $\frac{H}{D} = 3.1, 3.3$ and 3.5, the P–T coexisting regions terminate around $\frac{L}{D} \approx 1.83$; however, we have cut Fig. 3(c) at $\frac{L}{D} = 4.50 - 2.00 = 2.50$, for more complicated structures like two planar layers as well as one homeotropic can form at smaller rods.

By widening the slit pores ($H/D \geq 3.5$), at least three planar layers can exist even for very low densities because there is now enough space to form three planar layers. In other words, there is a special range of pore sizes ($3.5 \leq H/D \leq 4.0$) for the confined cylinders and parallelepipeds [51] in which the rods do not show any changes in the number of planar nematic layers.

Since the lowest number of planar layers increases for broader pores like $H/D = 8.0$ (Fig. 4 (d)), the lowest coexisting planar density with the homeotropic structure is shifted to a higher packing fraction due to increasing the number of allowable stable planar layers that can form in these pores. Hence, the planar structures are biaxial, and there is no intersection between the UP–BP and P–H transitions, *i.e.*, for $H/D \geq 3.5$, there is no direct transition from the UP structure to the H phase. In the confined parallelepipeds, this direct transition was seen just for very narrow pores where just two planar layers could form [51,24].

We do not observe a stable T structure for $\frac{H}{D} \geq 5$ within $H - 2D < L < H - D$. Therefore, this structure cannot survive for wider pores. This suggests that the hybrid planar-homeotropic T-type surface ordering is a distinct capillary effect that only occurs at small but sizeable window of pore widths. We reiterate that this structure is prohibited under conditions of extreme confinement $H < L$.

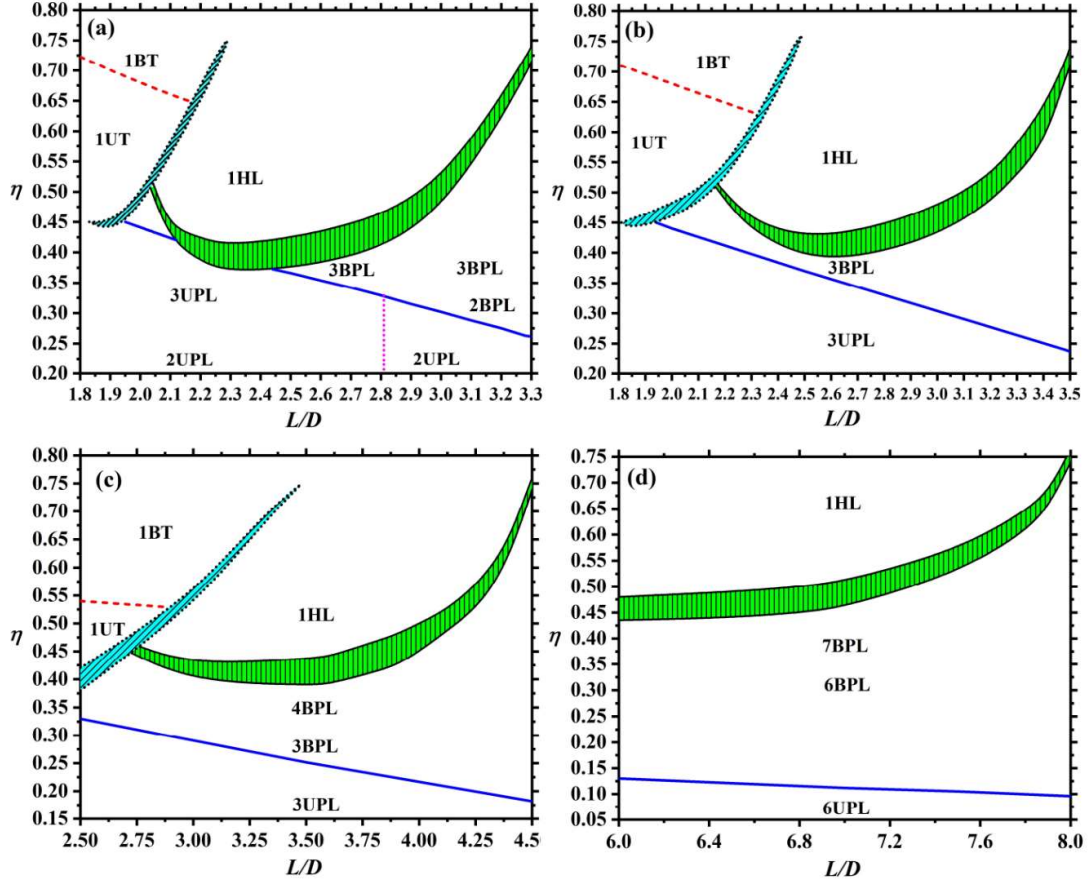

FIG. 4: (a) Phase diagram of $H/D = 3.3$; (b) $H/D = 3.5$; (c) $H/D = 4.5$; (d) $H/D = 8.0$.

For $H/D \leq 8.0$, the system forms the maximum allowable number of planar layers before crossing over into the homeotropic structure; however, the scenario is different in wider pores, where, interestingly, the homeotropic phase can form from the planar nematic with *n* layers as well as *n*+1 layers depending on the elongation of particles.

Fig. 5(a) demonstrates the phase diagram of cylinders for $H/D = 10.0$ and $9.00 \leq L/D < 10.00$. In this case, the following structural changes occur:

- for $9.00 \leq \frac{L}{D} < 9.50$:  7UPL $\xrightarrow{\text{second-order transition (blue solid line)}}$ 7BPL $\xrightarrow{\text{gradually}}$ 8BPL
  $\xrightarrow{\text{first-order transition (cyan region and patterned with vertical lines)}}$ 1HL

- for $9.50 \leq L/D < 10.00$:
  7UPL $\xrightarrow{\text{second-order transition (blue solid line)}}$ 7BPL $\xrightarrow{\text{gradually}}$ 8BPL
  $\xrightarrow{\text{first-order transition (pink region and patterned with tilted lines)}}$ 9BPL
  $\xrightarrow{\text{first-order transition (cyan region and patterned with vertical lines)}}$ 1HL

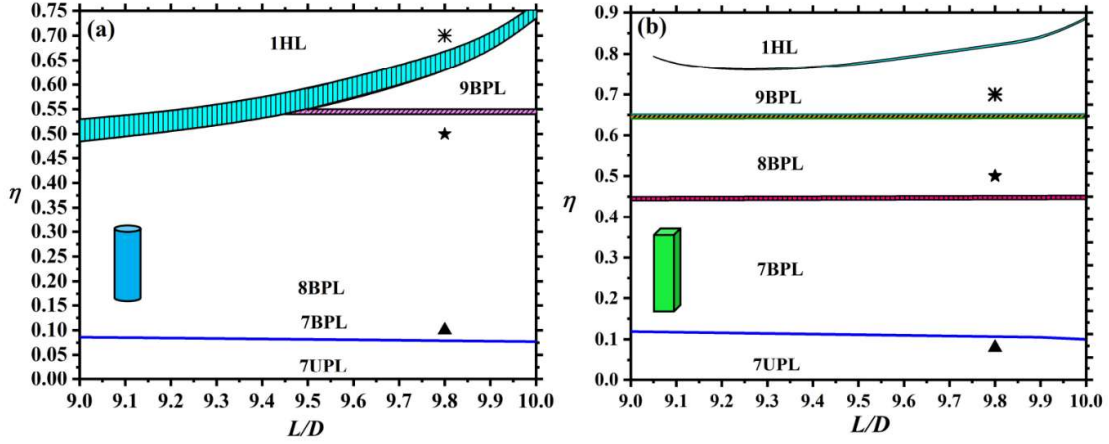

FIG .5: Phase diagram of $H/D = 10.0$ when $9.00 \leq L/D < 10.00$ of (a) cylindrical rods confined between two parallel hard walls and (b) same for hard parallelepipeds.

Evidently, there is a first-order transition between the planar phase with *n* layers and the planar structure with *n*+1 layers for rods with $9.50 \leq L/D < 10$. To draw a comparison between the phase structures of confined hard cylinders and hard parallelepipeds with square cross section $D \times D$ and the length $L$, the phase diagram of parallelepipeds was recalculated for $H/D = 10.0$ and $9.00 \leq L/D < 10.00$ (for details, the reader is referred to a previous paper by us [50]) and then plotted in Fig. 5 (b) (pink region with vertical lines and orange region with tilted lines). Although the UP–BP transitions of both systems depend very weakly on the aspect ratios, there is a weak decrease with $L/D$ in both cases with the second-order transition for the cylinders happening at lower packing fraction. For a better comparison, the density profiles of both shapes were plotted with the same parameters in Fig. 6. In fact, Fig. 6 (a) and Fig. 6 (b) present density profiles at $(L/D, \eta) = (9.80, 0.10)$ of the cylinders and the parallelepipeds, respectively where the cylinders exhibit a biaxial planar structure but the parallelepipeds show a uniaxial planar phase. The insets indicate the part of in-plane density profiles where one can see the number of planar layers more clearly. The transformation from UP to BP is an in-plane transition, and the density near the wall is responsible for this transition. The higher peaks of the cylinders near the planes indicate that the adsorption is stronger; hence, surface ordering occurs at lower packing fractions compared the parallelepipeds.

Since the formation of a new layer pushes the existing layers towards the confining surfaces, which leads to less available space for the existing layers, this structural change is accompanied by an increase in the free energy value at some packing fractions, and a first-order phase transition takes place to join the old and the new states. This occurs between all two biaxial nematic planar phases of parallelepipeds where the planar adsorption of these particles is weaker at the walls (compare Figs. 6 (c) and (d) at $(L/D, \eta) = (9.80, 0.50)$). Moreover, the layers between the sharp peaks near the walls are as strong as those at the walls. Therefore, the number of planar layers changes discontinuously from 7 to 8 around $\eta = 0.45$ and 8 to 9 around $\eta = 0.65$ in both cases. The formation of a new layer in the middle of the pore is easier for cylinders, because the layers between the surface layers are only weakly developed. Hence, a new layer can form gradually without any transition, even at lower densities where the system reaches 8 layers. According to the insets of Figs. 6 (a) and (b), the layer in the middle of the pore is thicker for cylinders, and the layers around it are very weak. Therefore, it can be divided into two layers easily, whereas the planar structure with 8 layers forms continuously from 7 layers around $\eta = 0.20$. In other words, the excluded volume gain is higher in the case of parallelepipeds than in the case of cylinders. As discussed earlier, the translational entropy (ideal gas contribution) competes with the excluded volume in the layering transition. The translational entropy is the same for both shapes; however, the excluded volume is different. Hence, the higher excluded volume gain is responsible for these results.

As the packing fraction increases, the 8BPL structure changes to the homeotropic structure for cylinders with elongation $9.00 \leq L/D < 9.50$ and to the 9BPL structure for $9.50 \leq L/D < 10.00$ with discontinuous transitions. The first one is a strong transition in which the coexisting region is wide, whereas the second one is a weak transition. Since the cylinders are very anisotropic for $9.50 \leq L/D < 10.00$, the H phase forms at very high densities beyond the range where 9BPL appears. The planar phase with 9 biaxial planar layers is more favorable for parallelepipeds because only the first-order 8BPL–9BPL transition occurs for $9.00 \leq L/D < 10.00$. This kind of transition depends weakly on $L/D$ and decreases very slowly by decreasing $L/D$ due to the easier rearrangement of particles. Since the 9BPL structure is very stable for parallelepipeds, a very weak 9BPL-1HL first-order transition occurs at very high densities ($\eta > 0.75$) with a critical point at $L/D \simeq 9.10$. For $\frac{L}{D} < 9.10$, the structure changes continuously from a planar to a homeotropic configuration. By increasing the intermediate regime density, almost all of the parallelepipeds align with the *x*-axis, as can be seen from Fig. 6 (c) and (d) where $\rho_y \simeq 0$ and $\rho_z \simeq 0$. For small enough particles (where there is no 9BPL-H transition), the particles gradually rotate away from the plane parallel to the wall and a homeotropic layer can be formed within the pore without any phase transition. Note that the homeotropic phase is a uniaxial one because by increasing the density most of the particles align along the *z*-axis, hence the in-plane translational entropy increases. Figs. 6 (e) and (f) demonstrate that the structure is 1HL for cylinders and 9BPL for parallelepipeds at $(L/D, \eta) = (9.80, 0.70)$, respectively.

The comparative Figs. 5 (a) and (b) convincingly show that cylinders exhibit a qualitatively different surface phase behavior than parallelepipeds. Most strikingly, the onset of P to H surface anchoring occurs at much more experiment-friendly packing fractions which makes the phenomenon more realistic and accessible experimentally.

Similar results can be found in Fig. 7 (a) for $H/D = 12.0$ and the cylinders with elongation $11.00 \leq L/D < 12.00$. The first-order planar layering phase transition from 10 to 11 occurs at higher densities, since more layers can be stabilized in wider pores. In other words, the transition from *n* to (*n*+1) is shifted to very high densities ($\eta > 0.75$) at even wider pores where the phases are more likely to exhibit solid, crystalline features that are not taken into account in our present model. Figs. 7 (b)–(d) show the relevant density profiles at $(L/D, \eta) = (11.80, 0.57)$, $(11.80, 0.62)$, and $(11.80, 0.70)$ where the structures are 10BPL, 11BPL, and 1HL, respectively.

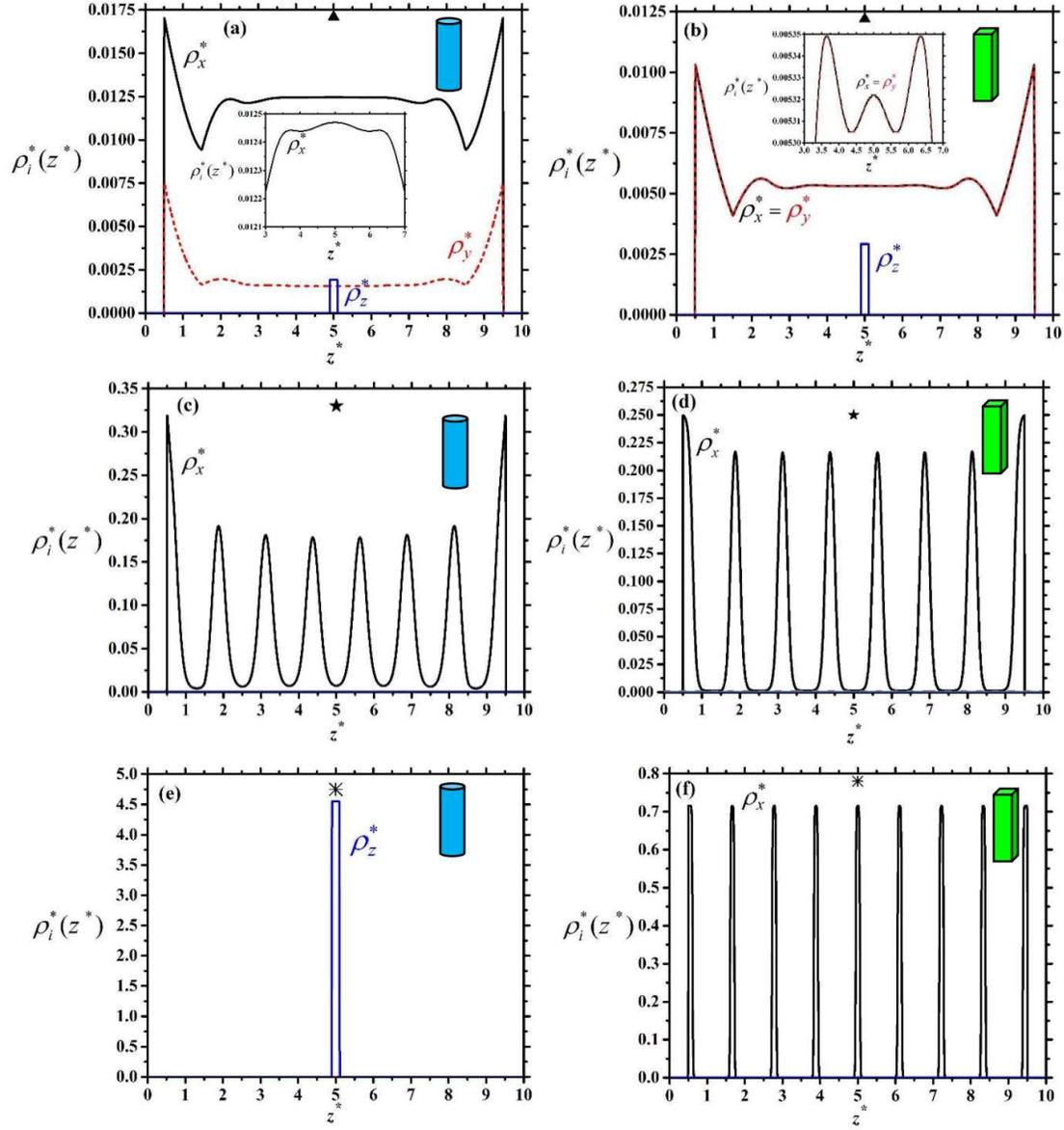

FIG. 6: Density profiles for cylinders and parallelepipeds confined in wide pores with $H/D = 10.0$ corresponding to the symbols in Fig. 5. (a) and (b) at $(L/D, \eta) = (9.80, 0.10)$; (c) and (d) at $(L/D, \eta) = (9.80, 0.50)$ (e) and (f) at $(L/D, \eta) = (9.80, 0.70)$.

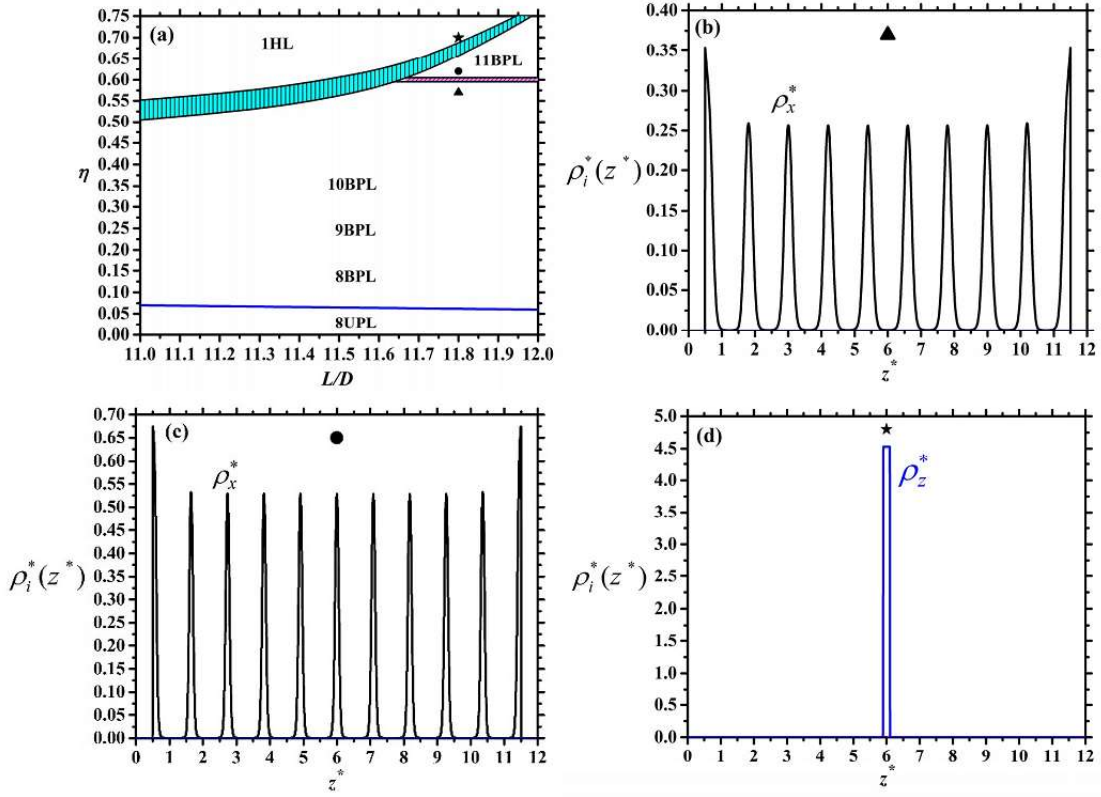

FIG. 7: (a) Phase diagram of $H/D = 12.0$ and $11.00 \leq L/D < 12.00$. The other panels indicate density profiles at (b) $(L/D, \eta) = (11.80, 0.57)$ where the phase is 10BPL (c) $(L/D, \eta) = (11.80, 0.62)$ and for 11BPL (d) $(L/D, \eta) = (11.80, 0.70)$, for 1HL. The pink area with tilted lines shows the coexisting region of the first-order 10BPL-11BPL transition and the cyan zone patterned with vertical lines presents the coexisting region of the $n$BPL-1HL state ($n = 10$ or $11$).

**Conclusion**

We have analyzed the effects of pore width and rod shape on the structural phases of hard cylinders confined between two parallel hard walls by using the Parsons–Lee theory in a restricted orientation model. Our focus is on the regime of strong confinement where strongly inhomogeneous structures appear near the walls and the system does not relax to its bulk structure even in the central region away the walls. All emerging structures are governed by strong interference between the opposing walls. One of the key results of our analysis is that the quasi-two-dimensional phase behavior of confined colloidal anisotropic particles is qualitatively different from observed the phase behavior observed in two and three spatial dimensions. For instance, in 3D (or at sufficiently large pore sizes) the isotropic-nematic phase transition is known to be first-order, whereas in 2D an isotropic phase emerges continuously from a nematic fluid upon decreasing particle density through a Kosterlitz-Thouless-type disclination unbinding transition [52, 53]. Therefore, examining rodlike particles confined in narrow pores can shed light on the quasi-2D regime where the rods are constrained to self-organize in a space between two and three spatial dimensions where capillary-induced ordering near the confining walls become prominent [54].

Planar ordering is the favourite anchoring symmetry at low and intermediate densities with strong adsorption occurring at the walls. At a critical packing fraction, purely repulsive extrinsic interactions induce a symmetry breaking transition from a uniaxial towards a biaxial planar structure which is generated by a trade-off between excluded volume entropy and orientational entropy as per

Onsager's original argument [15]. Similar results were observed in strictly 2D system for hard bodies confined between parallel hard lines, for example, ellipses, discorectangles, and rectangles [55-57]. The onset of biaxiality is accompanied by a second-order phase transition in the plane of the walls occurring at higher densities by decreasing the shape anisotropy due to a lower packing entropy gain with in-plane ordering. Biaxial order was observed in pore sizes of $H/D \geq 3.5$ while there is a special aspect ratio $(L/D)_s$ for $H/D = 3.1, 3.2$ and $3.3$ which for $H/D - 1 \leq L/D < (L/D)_s$, for which biaxiality does not occur. In addition, for $H/D = 3.1, 3.2$ and $3.3$, there are two kinds of uniaxial-biaxial phase transitions, namely one 3UPL–3BPL and 2UPL–2BPL. These two lines intersects at $L/D \approx 2.81$ in which the first kind of transition occurs for $L/D < 2.81$ because the weakly anisotropic particles are packed very well with the three-layer uniaxial planar state within the pore. Hence, this enables a transition toward the biaxial phase at higher densities. Such a behavior was also reported for confined hard cuboids with a square cross section [6,51] where there is a first-order layering transition from *n* to *n*+1 planar layers as opposed to the present case of cylinders. Furthermore, biaxial order was reported for confined rod-shaped particles such as ellipsoids [58], spherocylinders [34], and the stability of biaxial order is not sensitive to the cross section of rods.

Unlike parallelepipeds, in sufficiently wide pores and at some aspect ratios, the homeotropic structure is more stable than the planar nematic with *n*+1 layers. This finding indicates that the shape of the cross section of the particles seriously affects the planar nematic layering transitions and the stable phases that emerge.

Our results further suggest that there are always two planar layers at the walls; however, a homeotropically ordered layer competes with the planar one in the middle of the pore. Therefore, the planar phase changes to the homeotropic one for all cases studied as the packing fraction increased. The transition can be rationalized from a reduction of the packing entropy associated with perpendicular ordering at high packing fractions. We find that this structure is more stable for cylinders than for parallelepipeds because the cuboids are easier to accommodate in a planar phase with *n*+1 layers, and the H phase only becomes stable at an unphysically high packing fraction of parallelepipeds [51]. For cylinders, we find an absence of layering phase transitions from *n* to *n*+1 layers (such a behavior has been recently reported in Ref.[59] for confined hard ellipsoids studied by density functional theory) except in sufficiently wide pores and very anisotropic rods where two kinds of transitions from P to H occur in the pore *i.e.*, a nBPL–1HL transition and a (n+1)BPL–1HL one. The first-order planar layering transitions were reported for all cases studied in Ref. [50]. Hence, the width of the pore and the shape anisotropy of the cylinders are key factors in the stabilization of (*n*+1)-layered planar nematic structures confined within pores. The observation of a first-order nematic layering phase transition could challenge some previous findings [25]. However, a P–H transition was also reported for stiff-polymer rings [60] and also evidenced from changing the wall penetrability in a system of hard Gaussian overlap particles [61]. It has been reported in certain thermotropic systems [62]. Unreported in previous cases, however, we provide evidence of the formation of T structure emerging from a the homeotropic monolayer of confined hard cylinders for $\frac{H}{D} < 5$ which is stable in a sizeable pocket in the phase diagram. This phase can also form from the planar state for small enough rods and develops biaxial order upon increasing the density. We argue that the formation of T structure is a capillary effect since it only survives in sufficiently narrow pores of intermediate width, while vanishing if the pores become too wide (3D bulk limit) or too small (2D bulk limit). We also provide evidence for a concentration-driven reentrant uniaxial-biaxial (U-B-U-B) phase sequence in a wide region of intermediate pore sizes. Most importantly, we demonstrate that the circular cross-section of the rods plays a very important role in stabilizing the planar-to-homeotropic transition towards a much more accessible packing fraction, while at the same time rendering the transition much more strongly first-order than observed for parallelepipeds.

Moreover, since the PL theory is not capable of treating solid phases, we did not analyze possible crystalline in-plane and out-of-plane phases. Thus, the confined fluid may freeze first and then transform into a novel structure. On the other hand, the complex phases, for instance with in-plane crystallinity, could occur for large packing fractions. Intuitively, we cannot preclude that any system with a packing fraction higher than 40-50 % has a tendency to crystallize. Note that, in practice, crystal phases are usually strongly suppressed by length dispersity of the rods which is very prominent in many experimental rod systems [63,64].

On the theoretical side, and interesting perspective could be to analyze the present model within fundamental measure theory (FMT), which has been developed for freely rotating non-spherical hard particles [65-67] and hard prolate and oblate parallelepipeds using the Zwanzig approximation [68]. Note the T-configuration of particles may be stabilized by the restricted-orientation constraint therefore some cautions should be taken regarding the stability of the confined T phase when the restrictions in orientations are removed. However, we have strong evidence to believe that our results are qualitatively correct because the presence of the layering phenomena was demonstrated in 2D simulations of confined freely rotating squares and rectangles between two hard lines [55,69] and in 3D confined thin films [25, 27].

## Acknowledgment

R. A. thanks Prof. Szabolcs Varga and Prof. Peter Gurin for fruitful discussions.